\documentclass[a4paper]{article} 
\pdfoutput=1
\usepackage{graphicx}
\usepackage{jheppub}
\usepackage{slashed}
\usepackage{multirow}
\usepackage{multicol}
\graphicspath{{figures/}}

\newcommand{\changed}[1]{{#1}}

\newcommand{\mttwo}{M_{T2}}
\newcommand{\hide}[1]{}

\newcommand{\pxmiss}{{{\slashed{p}_x}}}
\newcommand{\pymiss}{{{\slashed{p}_y}}}
\newcommand{\vecPtmiss}{{\vec{\slashed{p}}}}

\newcommand{\vecv}{{\vec v}}
\newcommand{\vecp}{{\vec p}}
\newcommand{\vecq}{{\vec q}}

\newcommand{\vecs}{{\vec s}}
\newcommand{\vect}{{\vec t}}

\notoc 
\begin{document}
\title{\changed{Bisection-based asymmetric $\mttwo$ computation:\\{\large A higher precision calculator than existing symmetric methods}}}
\author[a]{Christopher~G.~Lester}
\affiliation[a]{Department of Physics, University of Cambridge}
\emailAdd{lester@hep.phy.cam.ac.uk}

\author[b]{and Benjamin Nachman}
\affiliation[b]{Department of Physics, Stanford University}
\emailAdd{bnachman@cern.ch}


\abstract{
An $\mttwo$ calculation algorithm is described.  It is shown to achieve better precision than the fastest and most popular existing \changed{bisection-based} methods.  
\changed{Most importantly, it is also the first algorithm to be able to reliably calculate {\bf asymmetric} $\mttwo$ to machine-precision,  at speeds comparable to the fastest commonly used symmetric calculators.}}

\maketitle

\section{Introduction}
For the purposes of this document, we will define the kinematic variable $\mttwo$ in the following way:
\begin{align}
  \mttwo(m_s,\vecs,m_t,\vect,\vecPtmiss; \chi_1, \chi_2)=\min_{
  \begin{array}{c}
\vecp,\vecq \mbox{ s.t. }\\
\vecp+\vecq=\vecPtmiss
\end{array}
  }\left\{\max\left[
      M_T(m_s, \vecs, \chi_1, \vecp), 
M_T(m_t, \vect, \chi_2, \vecq) 
\right]\right\}\label{eq:maindef}
\end{align}
where the {\it transverse mass} is given by
\begin{align*}
  M_T(m,\vecv, \chi, \vecp) = \sqrt{m^2 + \chi^2 + 2 \sqrt{m^2+|\vecv|^2}\sqrt{\chi^2 + |\vecp|^2} - 2 \vecv \cdot \vecp },
\end{align*}
\begin{multicols}{2}
\noindent in which
$\vecs$, $\vect$, $\vecp$, $\vecq$ and $\vecPtmiss$ are all real two-vectors, and the remaining quantities are real scalars which may all be assumed to be non-negative as they only enter through their squares.  Until fairly recently, the majority of $\mttwo$ usage in experimental literature concerned itself only with the so-called `symmetric' case, $\chi_1=\chi_2$, which is also the form in which $\mttwo$ was first proposed \cite{Lester:1999tx}.  However, there is a growing interest in the `asymmetric' case, $\chi_1\ne\chi_2$, \cite{Barr:2009jv,Konar:2009qr}, which is a powerful tool for reducing asymmetric backgrounds to symmetric signal processes.  For example, it was recently suggested to use asymmetric $\mttwo$ to suppress the dominant two-lepton $t\bar{t}$ background in searches for the supersymmetric top quark partners decaying into one charged lepton~\cite{Bai:2012gs}.  Such searches tend to require a large transverse mass, $M_T>M_W$, from the selected charged lepton and the missing momentum from the undetected neutrino (neutralinos in the signal can make $M_T\gg M_W$).  This eliminates most of the one charged lepton $t\bar{t}$ as $M_T\leq M_W$ and what is left is a two lepton $t\bar{t}$ background with an unidentified electron/muon or hadronically decaying $\tau$ lepton.  The natural choices for $\chi_1$ and $\chi_2$ are then $M_\nu$ and $M_W$.  When the rest of the $t\bar{t}$ topology is correctly assigned, $\mttwo$ has an endpoint at $m_\text{top}$ for the background.  This property helped to extend the most stringent limits on $\tilde{t}\rightarrow t\tilde{\chi}_1^0$ production~\cite{Aad:2014kra}.  Asymmetric $\mttwo$ has been used to search for many more topologies, including cases in which the signal processes are expected to have low values of $\mttwo$ as in three body $\tilde{t}\rightarrow bW\tilde{\chi}_1^0$ decays.  For these cases, the fact that a Jacobian peak pushes probability near the kinematic maximum of the $\mttwo$ distribution is crucial.

In general, there is no analytic, closed form formula for $\mttwo$.  While it is a relatively trivial matter to construct an algorithm capable of evaluating $\mttwo$ by using an off-the-shelf numerical minimiser and the definition given above, such methods are usually very slow, since the minimum is often on a singular fold or crease, and error prone as the minimum can be near infinity, or the crease may be steep sided but have a very shallow gradient along the fold.  Therefore, ever since the creation of $\mttwo$, there has been a need for fast and accurate methods for performing the minimisation.  Many of the existing methods can be found collated in \cite{oxbridgeStransverseMassLibrary}.  The publication in 2008  of the `elliptic bisection' method  of Cheng and Han \cite{Cheng:2008hk} caused a minor revolution in the world of $\mttwo$ algorithms. \changed{One of the most exciting things about their implementation (which we will refer to as `CH') was that it was very deterministic, having a computation time $\tau$ proprtional to the desired number of decimal places of precision $D$ in the final answer, i.e.\ $\tau \propto D$.  Full details are given later, but it suffices here to note that this performance is achieved by, at all times, ensuring that the final answer ($\mttwo$) lies on an interval of the real line, an interval which can be repeatedly bisected while maintining its validity.  It is this predictability that makes the method useful.  Not only can the method, in principle, reach arbitrarily high precision given enough bisections, but the method is inherently `fast' as it does not, indeed cannot, make wrong turns and head down the same dead-ends that off-the-shelf minimisers find themselves in.}  The CH bisection-based implementation has since been regarded as the {\it de-facto} fastest and most reliable method of evaluating the symmetric form of the variable.  \changed{Indeed, throughout this document, we will often use the term `fast' for any methods that are  `bisection-based', understanding this to mean that the methods has $\tau \propto D$. Our own method will fall into this category.}

$\mttwo$ is, from a theoretical perspective, no harder to compute in the asymmetric case than in the symmetric.  However, as a result of the historical accident that no-one was interested in asymmetric $\mttwo$ in 2008, the classic CH algorithm\footnote{The CH implementation is downloadable from \cite{zenuhanStransverseMassLibrary} and is also distributed within \cite{oxbridgeStransverseMassLibrary}.} is only capable of evaluating symmetric $\mttwo$.  Consequently, physicists needing to evaluate asymmetric $\mttwo$ have had no option but to either: (1) attempt to compute the variable themselves with generic minimisers, exposing themselves to the dangers and severe performance penalties mentioned above,  or (2) attempt to modify the CH algorithm to cope with the asymmetric case.   Neither of these options have been achieved satisfactorily. Ample evidence that the latter route (CH modification) is much harder than one might naively expect may be found in Walker's Table $1$ \cite{Walker:2013uxa} which shows that CH modifications which do not pay enough attention to the many areas of numerical instability that are carefully worked around in the CH algorithm, can easily lead to errors of hundreds of GeV in $\mttwo$ evaluations.  It is possible to find ways around the majority of these issues, as  Walker \cite{Walker:2013uxa} demonstrates elegantly, but only at some considerable cost in algorithm complexity.\footnote{Note that \cite{Walker:2013uxa} provides perl implementations of algorithms that can reliably evaluate $\mttwo$ in both the symmetric and asymmetric forms considered herein (and in some others).  Though the choice of language makes these implementations slow, they would probably become much faster if re-written in a different language.  At present, being more than a factor of 1000 slower than CH, these implementations are not counted as `fast' herein.  Nonetheless, it will be interesting to see (1) how much faster they could be if reimplemented, and (2) whether they can also reach machine precision in all the cases where CH struggles, described later.}

Given the above, the primary goal of the work written up herein, is to produce a reliable  $\mttwo$ calculator that is also fast \changed{(i.e.~$\tau\propto D$)} and capable of serving the previously un-served needs of those wanting to use the asymmetric case.  Rather than attempt to adapt the CH implementation itself, it was thought better to return to first principles, and ask why previous attempts to adapt CH had been so unsuccessful, complex or error prone.

\section{The good and bad bits of bisection algorithms} 
A key insight within CH \cite{Cheng:2008hk} was this: if it were possible to determine relatively quickly whether a `trial' value $M$ were greater than or less than the actual value of $\mttwo$,  {\bf without actually computing the actual value of $\mttwo$}, then by repeatedly bisecting a finite interval known to enclose the true value, it would be possible to determine $\mttwo$ in a cost effective manner.\footnote{In the most naive implementation the size of the interval shrinks by a factor of two on each `iteration', and therefore to achieve $n$ significant figures of precision in the answer, approximately $\log_{2}(10^n)$ bisections would be required.  In practical terms, this means that machine-precision on a modern desktop would be achievable in $O(30)$ iterations.  Even faster convergence might be possible if one had access to not only binary information (e.g.~is $M$ bigger or smaller than $\mttwo$) but to analog information that might allow intelligent bisection at somewhere other than the mid-point of the interval. We will see later that this is sometimes possible.}   This is very important, and will be heavily relied upon in the new algorithm described herein.

A second important result of \cite{Cheng:2008hk}, and later works, was that the trial value, $M$, could indeed be compared to the true value of $\mttwo$, without knowing the value of the latter.  We will not describe the origin of this test in detail here as it is described elsewhere.\footnote{The best reference for this is \cite{Walker:2013uxa} which is complete in all respects.  An accessible description of the main features can be found in \cite{Cheng:2008hk}, though it contains little concrete coverage of any of the special cases which are either not discussed or are left as exercises for the reader.} However the main points we will need to understand are the following:
\label{sec:items}
\begin{enumerate}
  \item 
    A trial value $M$, together with $m_s$, $\vecs$ and $\chi_1$, defines  a (possibly empty) subset $R_1(M)$ of points within ${\mathbb R}^2$:

    \vspace{-5mm}
    $$ R_1(M)=\{\vec{p}\in\mathbb{R}^2:M_T(m_s,\vec{s}, \chi_1, \vecp) \le M\} $$
  \item 
    A trial value $M$, together with $m_t$, $\vect$, $\chi_2$ and $\vecPtmiss$, defines another (possibly empty) subset $R_2(M)$ of points within ${\mathbb R}^2$:

    	    \vspace{-5mm}
       $$ R_2(M)=\{\vec{p}\in\mathbb{R}^2:M_T(m_t,\vec{t}, \chi_2, \vecp) \le M\} $$
  \item
    Whenever $R_i(M)$ (for $i$=1 or 2) is non-empty, it comprises either (i) a set of points that are within or on the boundary of a non-degenerate ellipse, or (ii) a set of points that are within
    or on the boundary of a non-degenerate parabola,\footnote{We define a point in ${\mathbb R}^2$ to lie `within' a parabola if it lies on a line segment joining any two points on the parabola.}
    or (iii) a set of points lying on a straight `ray', starting at a point in ${\mathbb R}^2$ and radiating out to infinity in some direction (this being a limiting case of progressively narrower and narrower parabolas, under certain circumstances), or (iv) a single point (this being a limiting case of progressively smaller and smaller ellipses, under certain circumstances). 
  \item
    The two singular cases (iii) and (iv) described above occur if and only if $M$ is equal to the kinematic minimum of the $\mttwo$ distribution, i.e.~when $M=\max[m_s + \chi_1, m_t + \chi_2]$. \label{item:moo}
  \item
    $R_1(M)$ (respectively $R_2(M)$) switches from elliptical to parabolic character (or from point to ray, in the singular case) when the mass $m_s$ (respectively $m_t$) goes from non-zero to zero.\label{baditem}
  \item
    The value of $M$ is greater than or equal to the true value of $\mttwo$ if and only if $R_1(M)\cap R_2(M) \ne \emptyset$.\label{gooditem}
\end{enumerate}
Item \ref{gooditem} above is the key beneficial feature that allows the bisection algorithms to work, and is used in the new algorithm we propose.  Item \ref{baditem}, on the other hand, is the cause of most of the trouble in the existing implementations. It causes problems for three reasons which we will describe in turn: (a) co-ordinate instability,  (b) special case proliferation, 
and (c) Lorentz-vector choices. We will describe the effects in terms of $R_1(M)$ only, but the same arguments apply to $R_2(M)$.

\subsection{Co-ordinate instability}

As $m_s\rightarrow 0$, but before 0 is actually reached, the ellipses bounding $R_1(M)$ grow progressively longer and longer (and wider and wider at the same time) in a `one-sided' manner -- i.e.~one of the two ends of the ellipse stays more or less where it is, eventually forming the pointy part of the parabola, while the other end of the ellipse grows and swells to infinite length and width in order to `open up' into the infinite region that is the inside of the parabola.  Because of this growth, the coordinates of the centre of the ellipse, and the lengths of its semi-major and semi-minor axes, all tend to infinity.\footnote{Given that all inputs to $\mttwo$ are finite quantities, such growth to infinity can only arise from dividing by quantities that become close to zero.  It is astonishing how many such divisions can be found lurking in existing implementations.  The new algorithm proposed here, has only one non-constant division, and it is provably non-dangerous.  Much of the safety of the proposed implementation relies on this fact.}  Many existing implementations use these unstable co-ordinates internally, and have only limited protections against their being used in unstable regions.  For example, the CH method is aware of this potential problem, and so first compares the magnitude of $m_s$ to an ad-hoc threshold, and if $m_s$ is dangerously light, flips over to treating $m_s$ as if it were identically zero (even though it is not). In such cases the CH method is thus able to use a custom massless algorithm which, knowing only about parabolae, does not share the parametrisation problems of the general method.  The need to flip at an arbitrary point, however, limits the ability to calculate precise values for any value of $m_s$ below that threshold, and the precise positioning of the threshold might need considerable investigation and tuning, which complicates the development and use of the algorithm.
\label{sec:er}
\subsection{Special case proliferation}
\label{sec:um}
So far as the authors can tell, when testing for the intersection in item \ref{gooditem} of section~\ref{sec:items}, all existing implementations examine some sort of discriminant, or Sturm Sequence, which is able to make a statement about the number or character of the points which lie on the two conics {\bf {\underline{bounding}}} $R_1(M)$ and $R_2(M)$.   This is probably the largest single inefficiency such methods introduce.  The basic problem is this: if the boundaries of $R_1(M)$ and $R_2(M)$ intersect then it is clear that $R_1(M)\cap R_2(M)\ne\emptyset$ but the reverse is not true. For example, $R_1(M)\cap R_2(M)$ may be non-empty without intersection of their boundaries when, say, $R_1(M)$ is bounded by a small ellipse entirely contained within the interior of a larger ellipse bounding $R_2(M)$.  So if these discriminants determine that the boundaries do not intersect, additional special-case code then has to determine whether this is due to a trivial enclosure, or to non-intersection of the interiors.  Such special-case code is complicated by the fact that different tests may be needed for parabola-nested-within-parabola, ellipse-nested-within-parabola, ray-nested-within-ellipse, and ellipse-nested-within-ellipse!\label{sec:blah}
\subsection{Lorentz-vector implementations}
\label{sec:lv}
If the dangerous region occurs when $m_s$ is close to zero, but not exactly so,  why should we care? No one expects measurement of near-zero jet masses at a precision of 0.0001 GeV at the LHC.  Alas, many end users use computer libraries that store Lorentz-vectors internally in $(E, p_x, p_y, p_z)$ formats. If massless vectors are stored in such formats, and then subsequently the mass of the vector is requested, the packages will often return a number whose magnitude is $\sqrt{|E^2-p_x^2-p_y^2-p_z^2|}$, which can easily fail to be zero due to finite precision and rounding effects.\footnote{For example, the commonly used framework ROOT~\cite{root} has a method {\tt TLorentzVector::M()} which returns the product of ${\rm sgn}(E^2-p_x^2-p_y^2-p_z^2)$ and $\sqrt{|E^2-p_x^2-p_y^2-p_z^2|}$.}  When this magnitude is not zero, it is frequently very small, just where it can cause co-ordinate instabilities in $\mttwo$ algorithms, unless protected against.
\section{The new algorithm}
 
The principle innovation in the new algorithm is that it finds a new way of approaching and performing the intersection test in item \ref{gooditem} of section~\ref{sec:items}.   Whereas existing methods largely examine the {\em boundaries} of $R_1(M)$ and $R_2(M)$ for intersection leading to the complications described in Section~\ref{sec:blah}, the new method uses the direct test for the intersection of the {\em interiors} of conics described in \cite{Etayo:2006:NAC:1159669.1159671}.  Interestingly,  \cite{Etayo:2006:NAC:1159669.1159671} only claims that their test is applicable to testing for the overlaps of ellipse interiors.  { One of us (CGL) makes the conjecture that the  test in \cite{Etayo:2006:NAC:1159669.1159671}, without modification, is also valid for testing for the overlap of the interiors of a non-singular parabola with another non-singular parabola or with a non-singular ellipse. Two of the authors of \cite{Etayo:2006:NAC:1159669.1159671}  indicate  that this conjecture is likely to be true, and may provide a formal proof of this later if it is not already a known result (private communication).  While no mathematical proof is offered here, the validity of the conjecture to the extent to which it is needed to support the functioning of the new algorithm is experimentally verified by the computation of millions of $\mttwo$ values and their comparison to the existing fast, slow, and analytic computations (see Sec.~\ref{sec:validate}).}
 
The benefits of this test are not only that it is one method which is as suited to working with parabolae as with ellipses (or mixtures thereof) but that it also represents the conics in a matrix form whose coefficients can remain bounded even as quantities like $m_s\rightarrow 0$.  No special change in the representations of these conics occurs when $m_s$ or $m_t$ move in small amounts near zero.  It is these features that allows for easy machine-precision calculations of $\mttwo$, as no ad-hoc selection criteria (cuts) are needed to deal with any of the transitions described in Sections~\ref{sec:er}, \ref{sec:um} and \ref{sec:lv}.  

The second alteration is small but no less important, and concerns item~\ref{item:moo} of section~\ref{sec:items}. Since the one place that the test of \cite{Etayo:2006:NAC:1159669.1159671} can become singular is the one place where the trial has reached kinematic limit, or as close to this as the machine-precision will allow, we can use this as a stopping condition for the bisection, provided that the lower end of the initial bisection region sits on the kinematic limit.  In other words, the lower end of the kinematic limit is started, by fiat, on the kinematic minimum.  The upper limit is grown exponentially until it is found to be bigger than the true value of $\mttwo$.  Thereafter the intervals are bisected, updating the upper or lower endpoints, as appropriate.  If at any stage in this bisection process either of the conics looks singular, it will be known this can only have happened because the trial value $M$ has been forced down to the kinematic limit, or as close to that as the machine-precision will allow.  This is therefore a sign that the true value of $\mttwo$ either is at the kinematic limit, or is as close to it as the machine can determine.  So when such a condition is encountered, the singular nature of the conic terminates further evaluation and the kinematic minimum is returned.  
The advantage of this is that it allows a scale-free way of dealing with the very lowest $\mttwo$ values.

The third and final tweak in the new algorithm \changed{is of very little importance, and can be safely removed, if desired, without invalidating any important claim in this document.  If used, it provides a factor of three speed increase when evaluating $\mttwo$ values that lie at the kinematic minimum, but leads to a small 3\% slowdown in evaluations for other types of events.  Whether this feature is useful to users will depend on the populations of events they are working with.  We call it the `deci-section optimisation', \label{sec:decisection} and have enabled it by default, since most users will not notice its 3\% cost, and a small number of users may value the much larger speed increase it gives them. Examples of the small benefits gained by turning this optimisation on and off for a few plausble real-world secarios are shown later.}  Deci-sectioning works as follows: since any $\mttwo$ value returning the kinematic minimum will have produced only bisections which shrunk the bracketing interval in the high side, one can choose to bias bisections low (say to the bottom 10\% of the current bracketed range) until such time as a trial $M$ value is found to be {\em less} than the true value of $\mttwo$, after which bisections return to the centre of the bracketed interval.  By this means, answers near the kinematic minimum can be reached roughly three times ($\sim\log_2(10)$) more quickly than they would otherwise, with negligible loss of performance for answers in the `bulk'.

  \changed{Note that the important (first and second) innovations were primarily concerned with removal of numerical instabilities and special cases that increased the complexity or limited the performance of existing methods.
    Neither of these changes were needed to make the new method `fast' (i.e.~$\tau\propto D$) as `fast' comes for free with our use of a bisection-based method.  Nonetheless, the fact that both changes make our code smaller means that it is not unreasonable to expect that the cost-per-bisection for our algorithm should be of similar order or better than of the CH algorithm in the (symmetric) cases where both can be compared at identical precision.  A small amount of evidence supporting this suggestion is given in Table~\ref{tab:timing}, which shows timings for our algorithm when compared to CH at similar precision for some plausible event types.  Therein it is seen that our algorithm achieves the same precision as CH in about 50\% to 70\% of the CPU time.  Since our algorithm has a fixed cost per bisection, it should be noted that this performance gain should be independent of event samples and individual event configurations.  Turning on and off the deci-section optimsation  shows that it is marginally beneficial (at around the 5\% level) for these particular event samples, however, it is important to stress that such small differences will vary between  different compilers, compiler optimization levels, CPU architectures, and event kinematics.  The only important conclusion we wish to draw from these timing tests is that our algorithm's execution time, which  necessesarily scales like $\tau\propto D$, is not hobbled by any large constant of proportionaily compared to CH, and indeed appears likely to be faster than CH (at similar precision) under most circumstances. }

A single {\tt C++} header file fully implementing an $\mttwo$ calculator using on the above method is available as part of the first arXiv submission of this paper under `Ancillary files'.\footnote{See \url{https://arxiv.org/src/1411.4312/anc/lester_mt2_bisect.h}.}  Links to later versions or corrections (if any) will be signposted in the Oxbridge Kinetics library \cite{oxbridgeStransverseMassLibrary} and/or in the arXiv comment field.

\section{Validation}
\label{sec:validate}
\begin{figure*}
  \centering
  \includegraphics[width=0.45\textwidth]{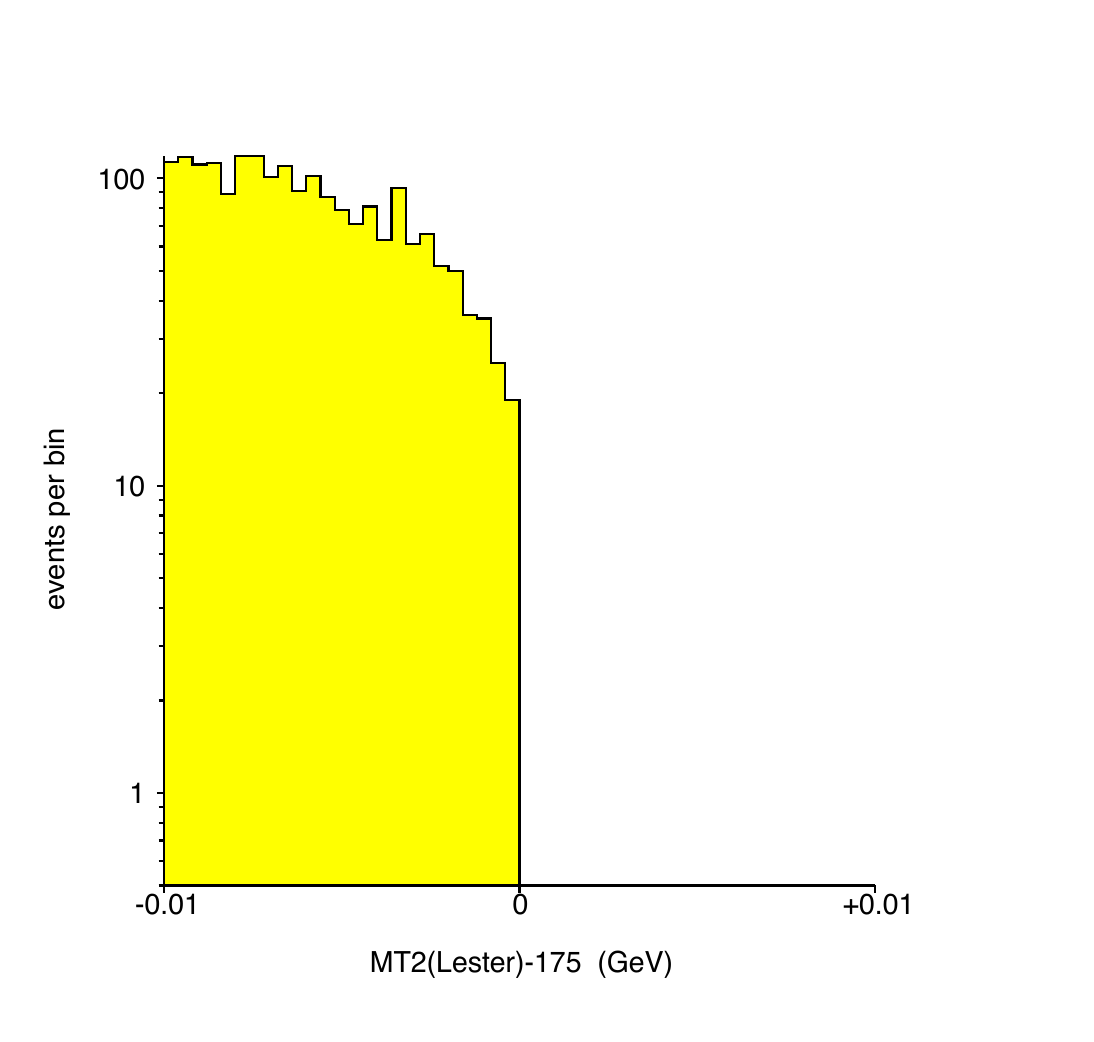}
  \includegraphics[width=0.45\textwidth]{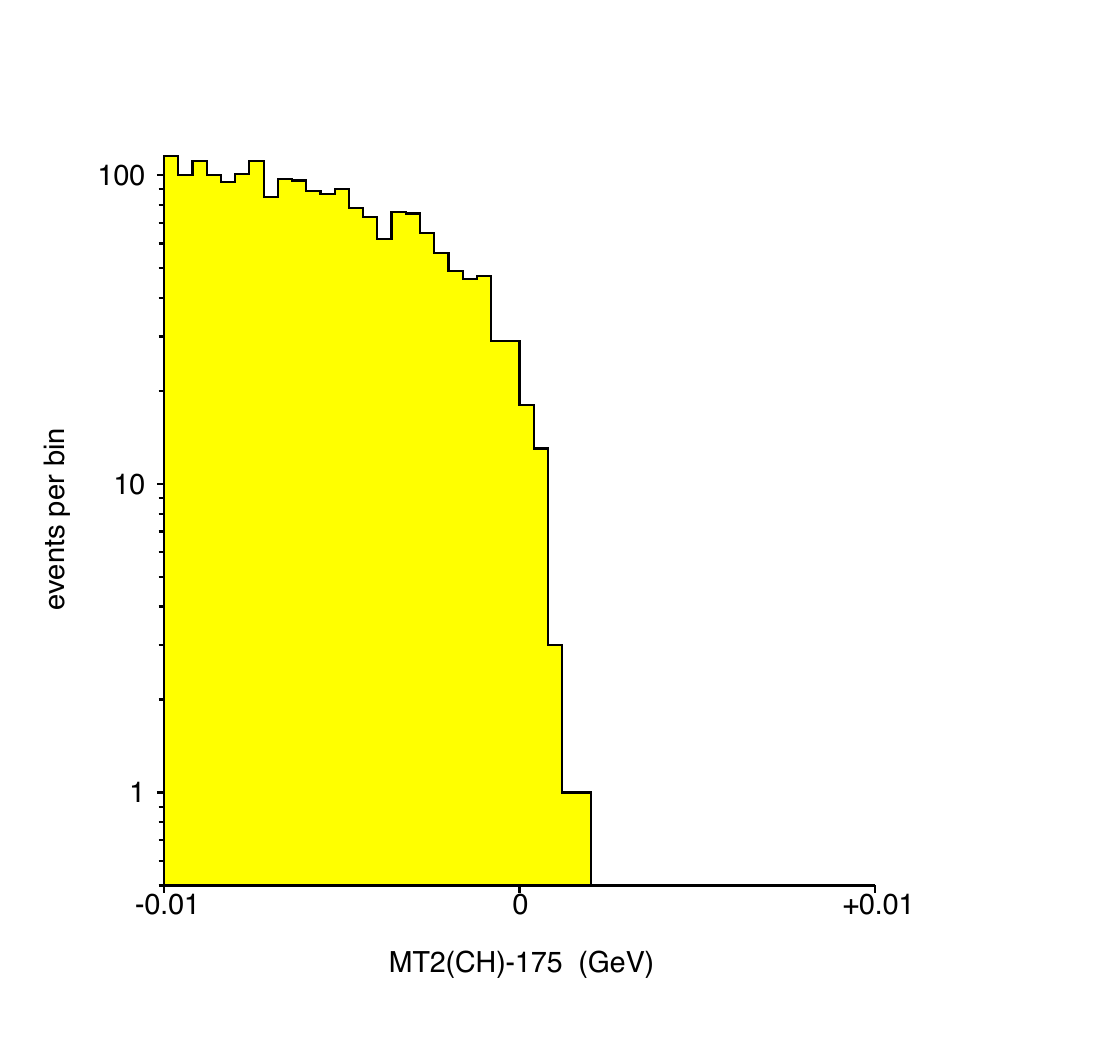}
  \caption{
    \label{fig:test}
    This figure shows a zoom in to the upper kinematic endpoint, $\pm  0.01$ GeV, of a $t\bar t$ $\mttwo$ distribution that ought to stop at $m_\text{top} = 175$ GeV (i.e.~at zero on the $x$-axes above).  It is noted that the new algorithm (left) stops at 0, as it should, while CH (right) is sometimes about 0.002 GeV too high.
}
\end{figure*}

\begin{figure*}
  \centering
  \includegraphics[width=0.45\textwidth]{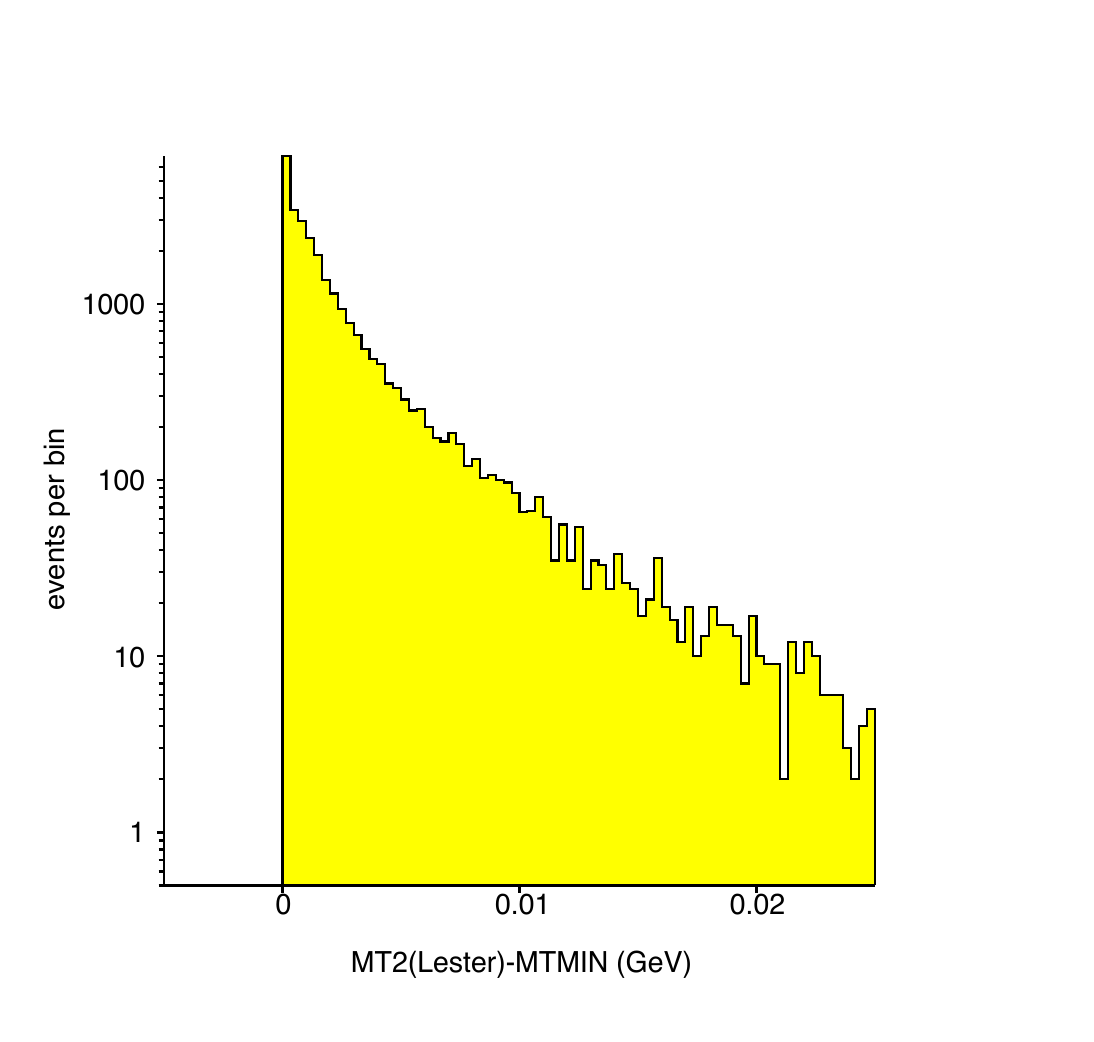}
  \includegraphics[width=0.45\textwidth]{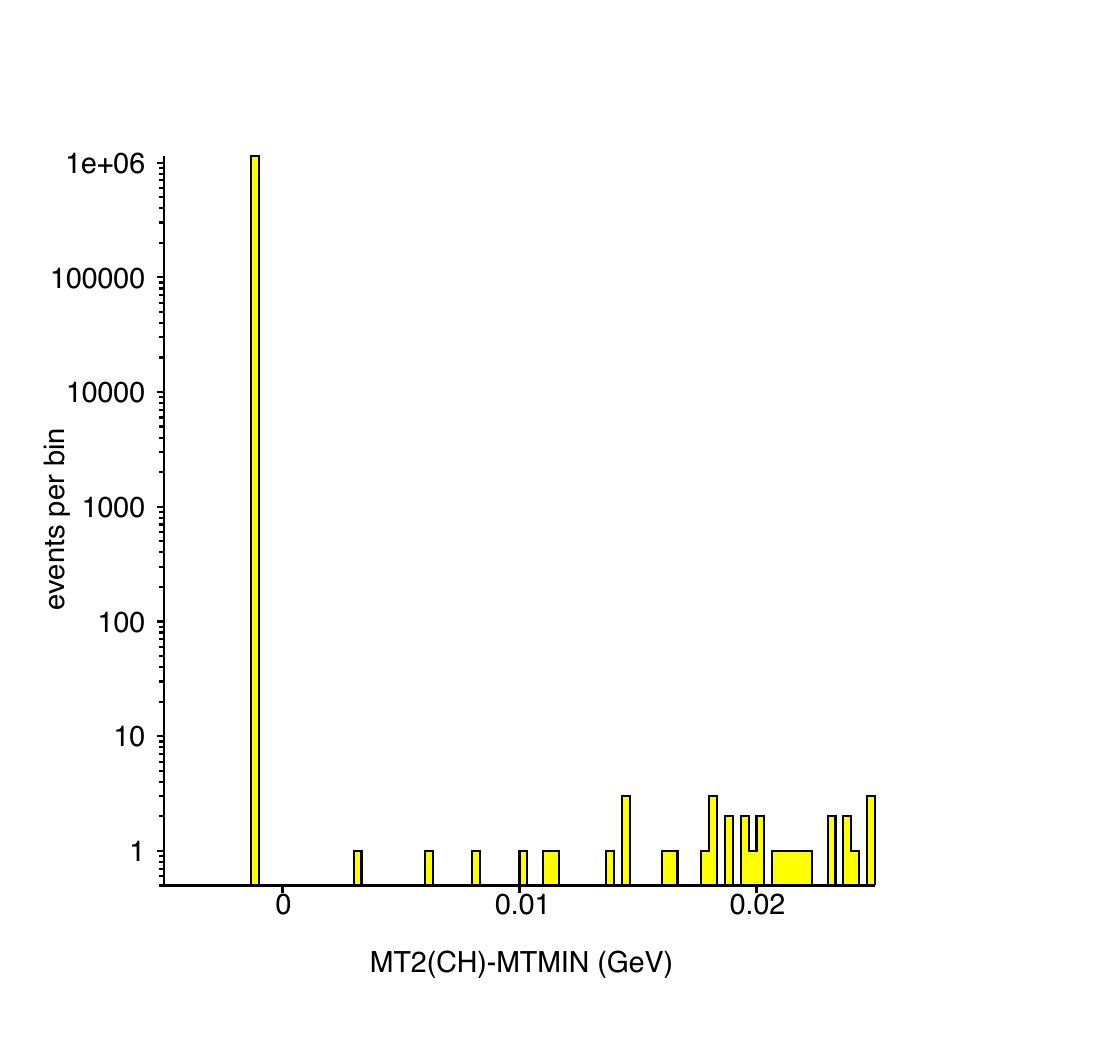}
  \caption{
    \label{fig:massgap}
    This figure shows the lower end of an $\mttwo$ distribution containing a high number of close-to-massless visible and invisible objects.  The kinematic minimum has been subtracted, so the distributions shown should never access negative positions on the horizontal axis.  The new algorithm (left) performs correctly, but the old CH algorithm (right) shows a large delta-function spike at an unphysical (negative) position, indicating that these $\mttwo$ values are lower than the kinematic minimum. An unphysical mass-gap is also observed in the CH case.
}
\end{figure*}

\begin{figure*}
  \centering
  \includegraphics[width=0.5\textwidth]{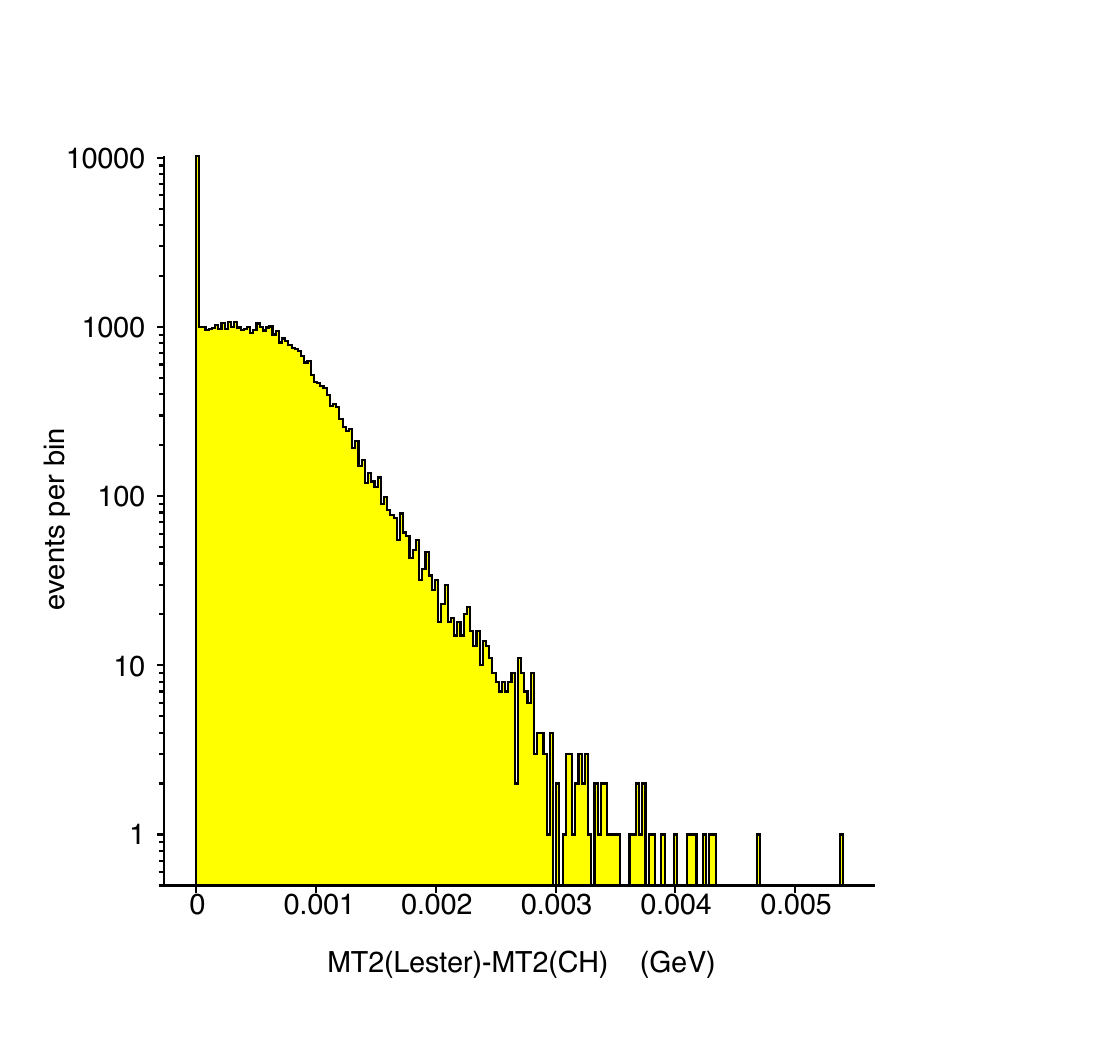}
  \caption{
    \label{fig:delta}
    This shows the difference between the value of $\mttwo$ calculated using the CH algorithm, and the value of $\mttwo$ calculated (for the same event) by our new algorithm.  The plot uses a realistic sample of $t\bar t$ events.  Leptons were all treated as massless, and $\chi$ was zero.  When supplying momenta to the algorithms,  the correct $l$+$b$ pair for each `side' of each event was used.  There is a narrow core of events that both algorithms get right, and a tail in  which (in this case) CH happens to be high at the 0.002 GeV level.  Some individual events from this tail were examined in detail (by inspecting the minimsation surfaces in arbitrary precision plotting programs and finding the location of the true minimum by hand) and the new answer was, in all cases, found to be the correct one.  An illustration of an investigation of this type (though not for one of these $t\bar t$ events) may be seen in Figure~\ref{fig:investigate}.
  }
\end{figure*}

\begin{figure*}
  \centering
  \includegraphics[width=0.5\textwidth]{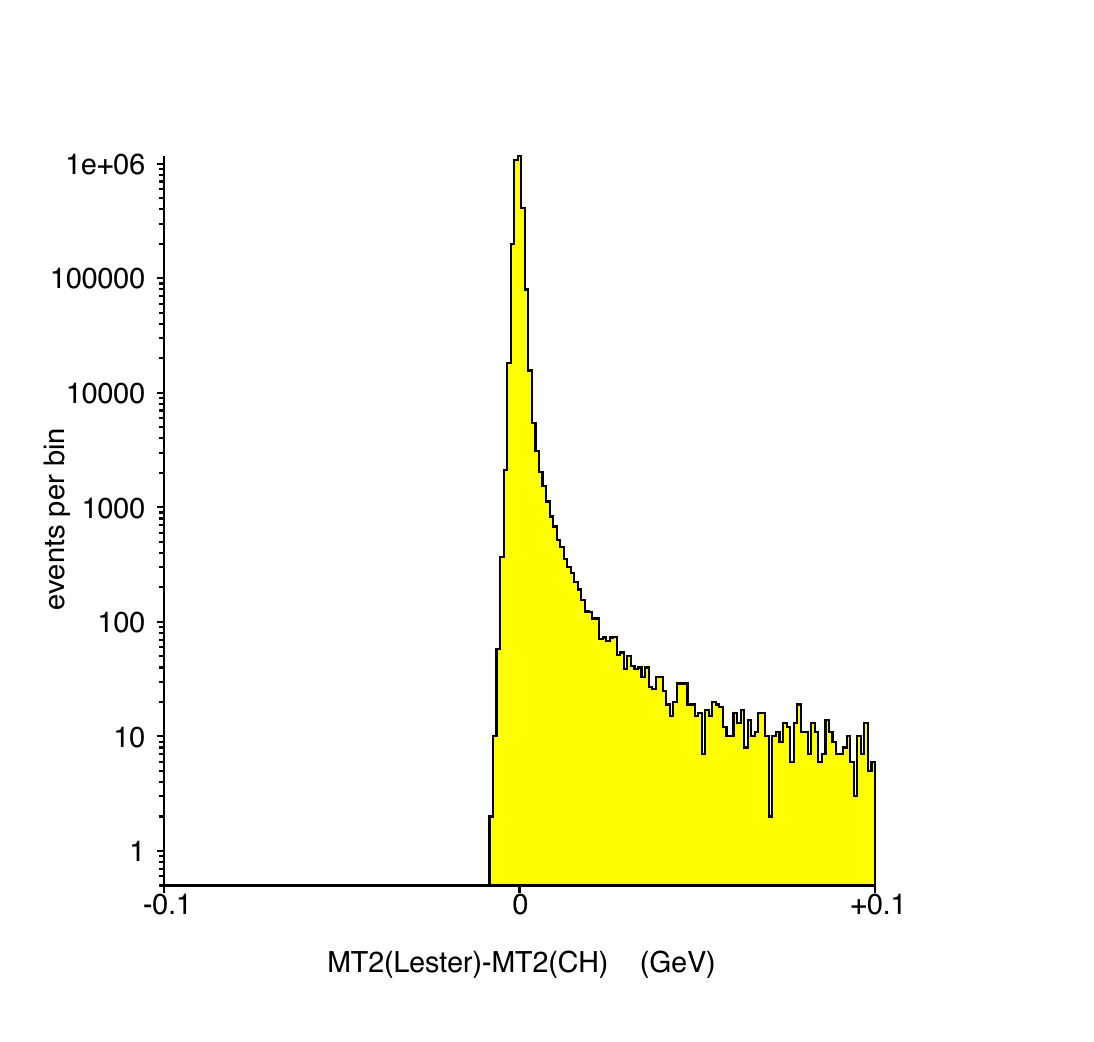}
  \caption{
    \label{fig:delta2}
    This shows the difference $\delta$ between the value of $\mttwo$ calculated using the CH algorithm, and the value of $\mttwo$ calculated (for the same event) by our new algorithm.  The plot uses a large population of somewhat 
    artificially constructed events, described below, designed to better illustrate the failure modes of existing algorithms seen in Figure~\ref{fig:delta}.  As in that figure, a central core of events for which both algorithms get similar $\mttwo$ values can be seen, but now that core is much broader, and there is a much larger tail in  which it is evident that the CH algorithm produces $\mttwo$ values that are considerably lower than ours.
    The events used here all begin with a pair of 300 GeV sleptons produced approximately 200 GeV above threshold in association with a recoiling object, such as a jet.  Each slepton is allowed to decay to a lepton and a neutralino.  The masses of the four daughter particles (two leptons and two neutralinos) are randomized in each event.  Each such daughter has a 50\% chance of being massless, and a 50\% chance of having its mass uniformly distributed between 0 and 10 GeV, independently of the masses of the other daughters in that event.  $\chi$ is always taken to be zero.  This is therefore a sample of events which frequently has a mixture of mass scales (massive and massless) in the inputs to  $\mttwo$.
    Some individual events from this tail were examined in detail (by inspecting the minimsation surfaces in arbitrary precision plotting programs and finding the location of the true minimum by hand) and the new answer is, in all cases, found to be the correct one.
    Such an examination, for the first generated event with a value of $\delta$ lying in the range $[0.09,0.10]$, may be seen in Figure~\ref{fig:investigate}.
  }
\end{figure*}

\begin{figure*}
  \centering
  \includegraphics[width=0.45\textwidth]{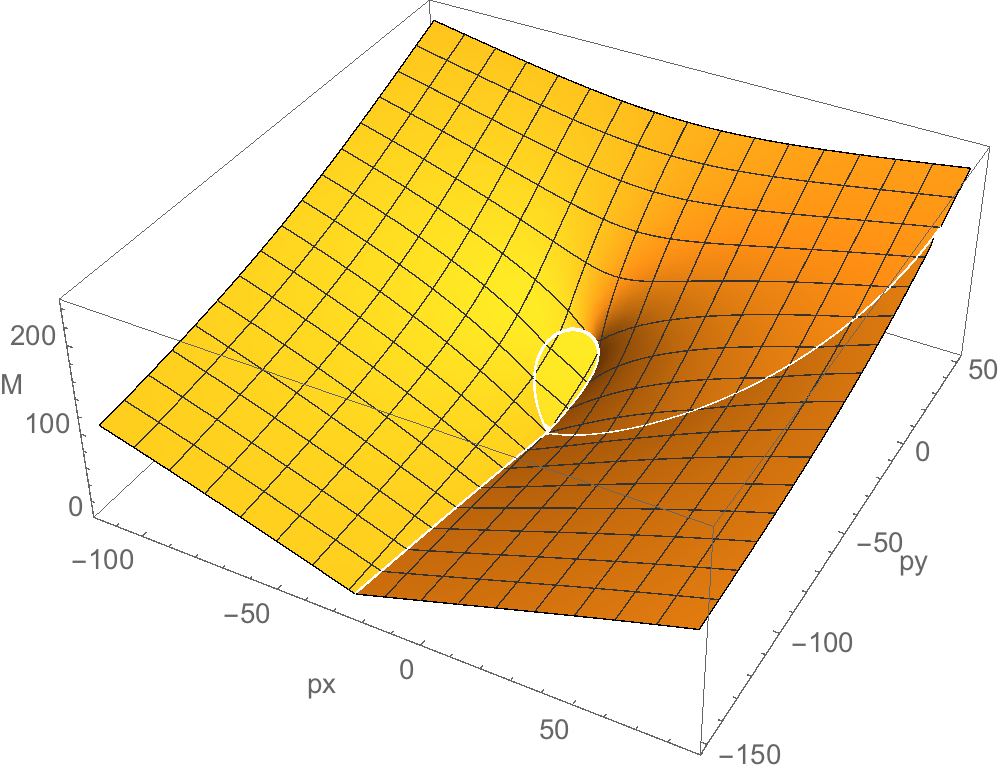}
  \includegraphics[width=0.45\textwidth]{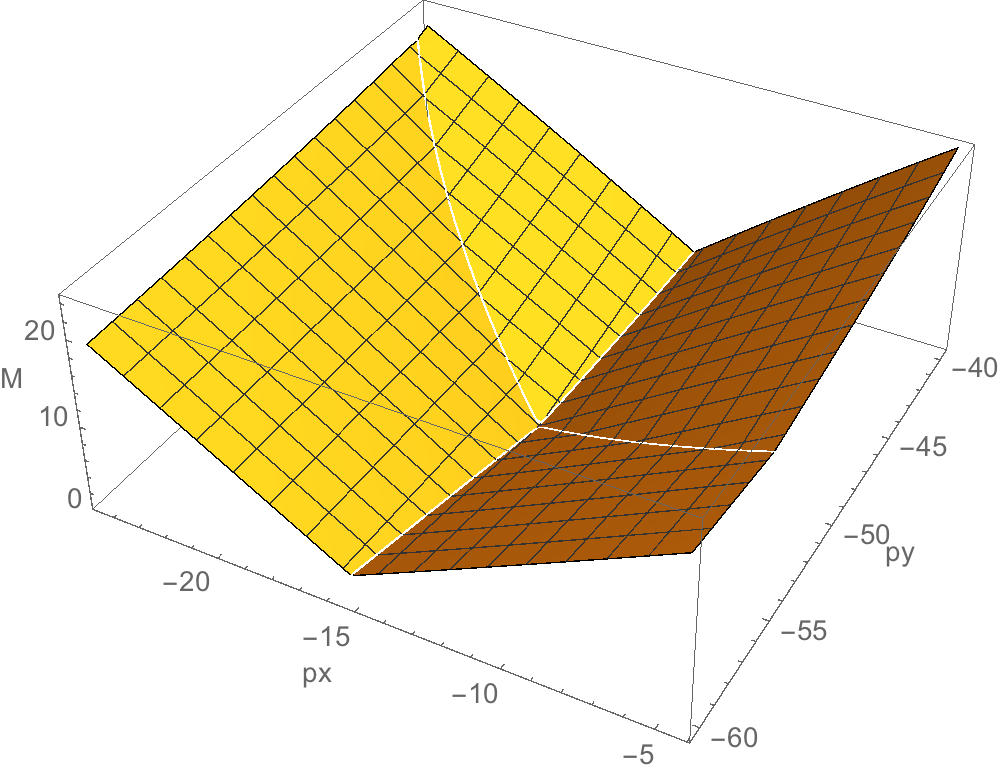} \\
  \includegraphics[width=0.45\textwidth]{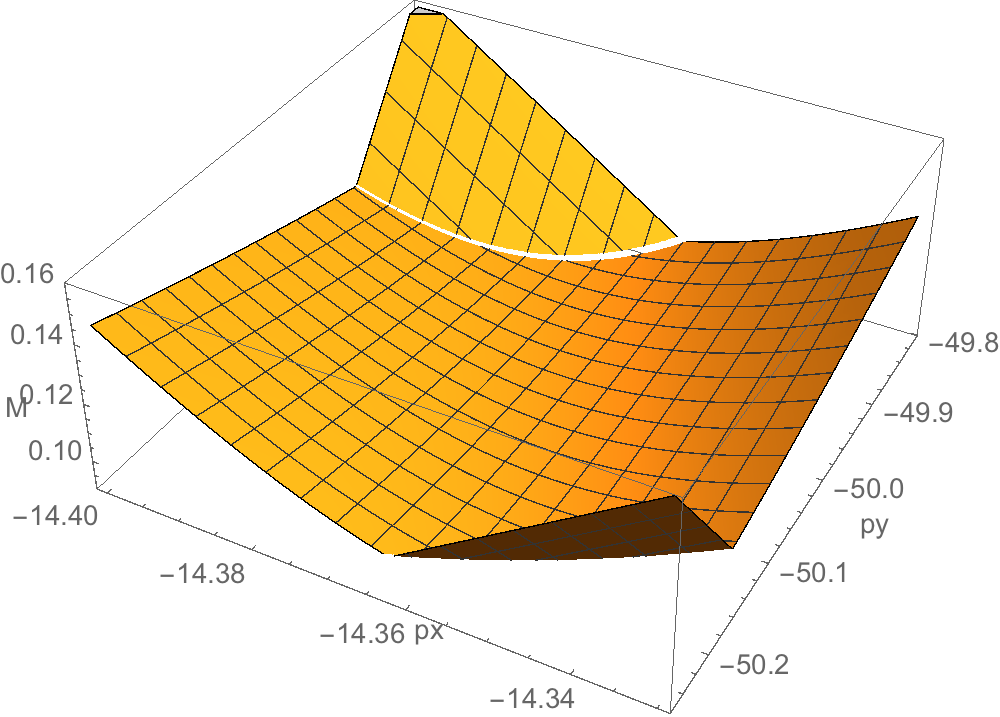}
  \includegraphics[width=0.45\textwidth]{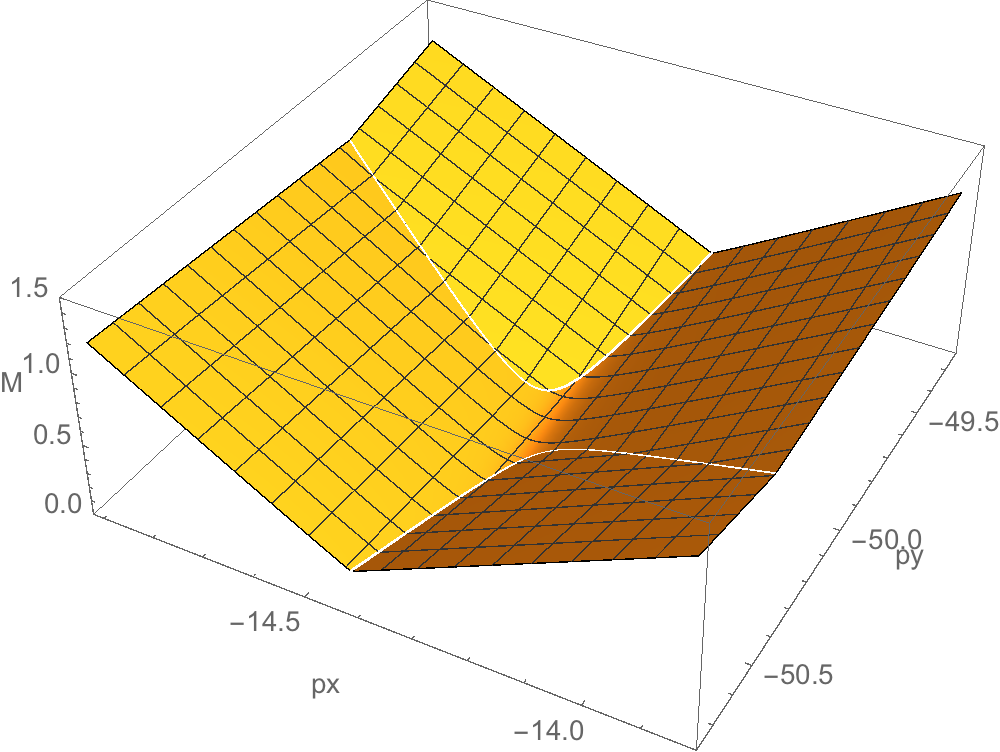}
  \caption{
    \label{fig:investigate}
    These plots, clockwise from top left, show progressively zoomed-in images of the surface $M = \max\left[
            M_T(m_s, \vecs, \chi_1, \vecp), 
            M_T(m_t, \vect, \chi_2, \vecq)
          \right]$, viewed as a function of the $x$ and $y$ co-ordinates of $\vecp$ (since $\vecq$'s coordinates are defined in terms of those together with the $\vecPtmiss$ constraint), drawn for the kinematic configuration of the event described in the caption of Figure~\ref{fig:delta2}. This event has 
         $m_s$     =    0, 
         $s_x$     =  -42.017340486, 
         $s_y$     = -146.365340528, 
         $m_t$     =    0.087252259, 
         $t_x$     =   -9.625614206, 
         $t_y$     =  145.757295514, 
         $\pxmiss$ =  -16.692279406, 
         $\pymiss$ =  -14.730240471, 
         and $\chi_1=\chi_2=\chi=0$.
         This surface is that over which the  minimisation in Equation~\ref{eq:maindef} proceeds, and so the smallest value of $M$ attained on this surface is the desired value of $\mttwo$ for that event.  It may be seen from the view at the largest scales (top left) that  the surface has a non trivial shape, with the minimum lying on what appears to be  the intersection of two folds.  Zooming in a little (top right), this assessment still appears to be true, but at higher zoom (bottom right) it is apparent that there is not an intersection here after all.  The real minimum lies at one of the two hyperbolic creases which are shown magnified in the highest zoom level (bottom left).  From further zooms (not shown) and inspection of the vertical ($M$) axis, it may be seen that the lowest point on this surface is indeed at the value of $\mttwo$=0.09719 returned by the new algorithm, and that this is inconsistent with the value returned by the CH algorithm (which was identically zero).  [All units in figure and its caption are in GeV.]
  }
\end{figure*}

\begin{figure*}
  \centering
  \includegraphics[width=0.45\textwidth]{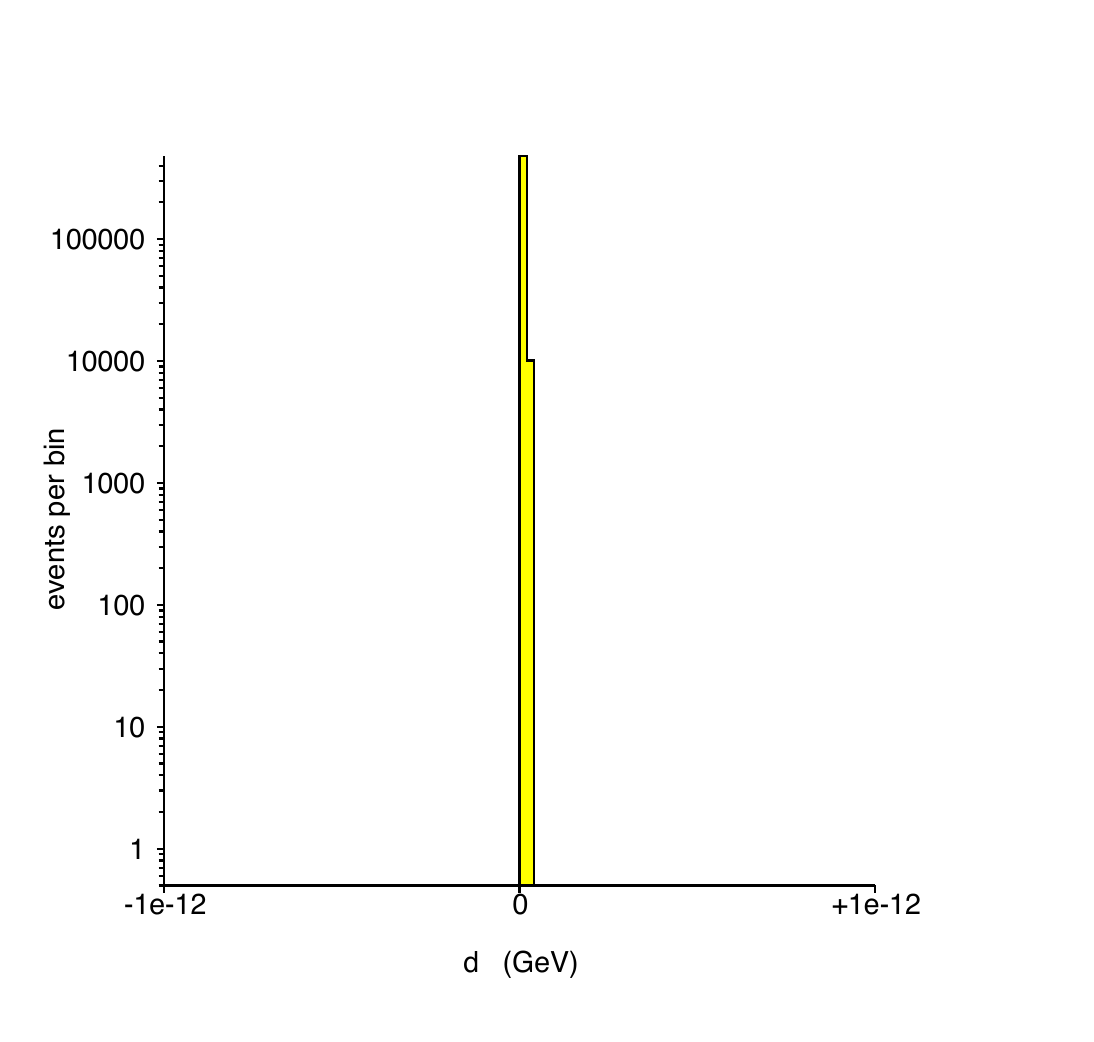}\includegraphics[width=0.45\textwidth]{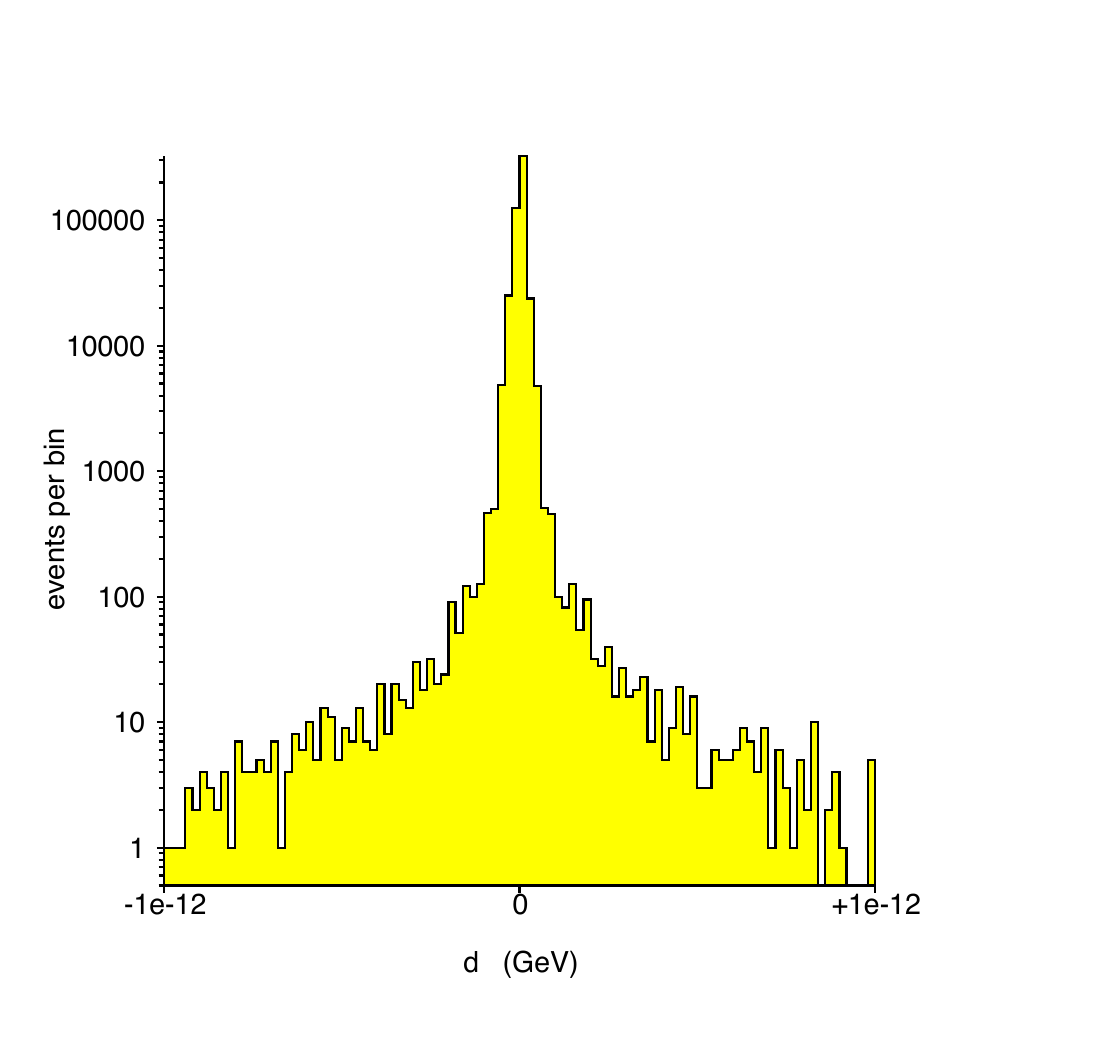}
  \caption{
    \label{fig:analytic}
    For each event of the type described in the text, the quantity $d$ plotted here is defined to be value of $\mttwo$ computed for that event using the new algorithm described in this paper, minus the value of $\mttwo$ computed using a known analytic formula.
    Left: A comparison between the new algorithm and the analytic calculation in the unbalanced case.  Right: the same comparison in the balanced case - now the analytic formulae has comparable calculation complexity as the numerical procedure. For definitions of balanced and unbalanced, see \cite{Cho:2007dh}. Each histogram contains 100 equally sized bins.  
  } 
\end{figure*}

A selection of validation plots are provided here, that selectively highlight the biggest areas of difference between the new algorithm and the CH baseline.   All tests were executed with a toy MC that pair produces on-shell particles.  In such cases, there are analytic formulae for the kinematic minimum and maximum of the $\mttwo$ distribution.  Figure~\ref{fig:test} illustrates the CH algorithm failing to accurately describe the upper end of a $t\bar t$ symmetric $\mttwo$ distribution at the 0.002 GeV level.  Figure~\ref{fig:massgap} illustrates the CH algorithm failing to cope with large numbers of near massless particles near the lower kinematic endpoint of another symmetric $\mttwo$ distribution.  Figure~\ref{fig:delta} is an example of a comparison in the bulk for a large sample of events containing a pair of top quarks.  \changed{Figure~\ref{fig:delta2} shows a similar comparison but this time for a sample of events which CH finds much harder to handle.  Investigations into why our algorithm sometimes produces different answers to CH have been conducted, by inspecting events in the tails of these distributions of differences.  An example of one such investigation is given in Figure~\ref{fig:investigate}.  All such investigations peformed by the authors to date have all indicated that the new algorithm produces the better answer when the different calculators disagree.} While the comparisons in the bulk of the distribution illustrate relative differences between the algorithms, it would be ideal if the new algorithm could be compared not only to CH but to known true values of $\mttwo$.  This is possible when the recoil (`upstream') momentum of the system of inputs to $\mttwo$ is zero - in this case there are known analytic formulae for asymmetric $\mttwo$~\cite{Konar:2009qr}.  Figure~\ref{fig:analytic} compares the new algorithm with the analytic formulae in two $t\bar{t}$ events for which one of the leptons is lost and so $\chi_1=m_\nu$, $\chi_2=m_W$, and the spread of $\mttwo$ is $\sim 100$ GeV.  The agreement between the machine precision numerical algorithm and the analytic calculation agree to better than $10^{-12}$ GeV, or one part in $\sim 10^{14}$.

\section{Summary}
Thus far the new algorithm has passed all validation tests it has been given, including many not decribed here. It can calculate both symmetric and asymmetric $\mttwo$ to machine-precision (one part in $\sim 10^{14}$ on the computer on which it was tested).  In cases where the new algorithm has been found to  produce different results to the de-facto standard calculator (CH),  the new algorithm is found to be correct.  Data in Table~\ref{tab:timing} indicates that, if asked to calculate to the same 0.002 GeV precision as the old CH algorithm, the new algorithm is almost two times faster and is capable of evaluating  $\mttwo$ values at $\sim$MHz rates.  If asked to calculate at machine-precision, the new algorithm  is at most 20\%-50\% slower than CH, and is still capable of 200 kHz evaluation rates.  There appears, therefore, to be many good reasons for trialling it a new reference implementation. 

\section{Acknowledgements}
CGL gratefully acknowledges the assistance of Gary Hughes and Mohcine Chraibi, authors of \cite{HughesMohcine}, for having pointed him towards \cite{Etayo:2006:NAC:1159669.1159671} after discussions about their paper, which would otherwise have formed the basis of this algorithm.  CGL also gratefully thanks Joel Walker and Ben Brunt for useful discussions, and Alan Barr for making many comments on the final draft.  Finally, CGL notes that three weeks after writing this algorithm, but about three weeks before writing up, he was contacted by colleague Colin Lally who, by chance, had independently come up with very much related ideas (sharing the matrix addition of conics discussed in \cite{Etayo:2006:NAC:1159669.1159671} as a starting point, for example).
CGL thanks Colin for suggesting ways in which the method described herein could potentially be improved. None of these ideas are yet implemented.

\bibliographystyle{JHEP-withSlacCitation}
\bibliography{paper}

\providecommand{\href}[2]{#2}\begingroup\raggedright\begin{thebibliography}{10}

\bibitem{Lester:1999tx}
C.~G. Lester and D.~J. Summers, {\it {Measuring masses of semiinvisibly
  decaying particles pair produced at hadron colliders}},  {\em Phys. Lett.}
  {\bf B463} (1999) 99--103,
  [\href{http://xxx.lanl.gov/abs/hep-ph/9906349}{{\tt hep-ph/9906349}}].

\bibitem{Barr:2009jv}
A.~J. Barr, B.~Gripaios, and C.~G. Lester, {\it {Transverse masses and
  kinematic constraints: from the boundary to the crease}},  {\em JHEP} {\bf
  11} (2009) 096, [\href{http://xxx.lanl.gov/abs/0908.3779}{{\tt
  arXiv:0908.3779}}].

\bibitem{Konar:2009qr}
P.~Konar, K.~Kong, K.~T. Matchev, and M.~Park, {\it {Dark Matter Particle
  Spectroscopy at the LHC: Generalizing M(T2) to Asymmetric Event Topologies}},
   {\em JHEP} {\bf 1004} (2010) 086,
  [\href{http://xxx.lanl.gov/abs/0911.4126}{{\tt arXiv:0911.4126}}].

\bibitem{Bai:2012gs}
Y.~Bai, H.-C. Cheng, J.~Gallicchio, and J.~Gu, {\it {Stop the Top Background of
  the Stop Search}},  {\em JHEP} {\bf 1207} (2012) 110,
  [\href{http://xxx.lanl.gov/abs/1203.4813}{{\tt arXiv:1203.4813}}].

\bibitem{Aad:2014kra}
{\bf ATLAS Collaboration} Collaboration, G.~Aad {\em et.~al.}, {\it {Search for
  top squark pair production in final states with one isolated lepton, jets,
  and missing transverse momentum in $\sqrt{s}=$ 8 TeV pp collisions with the
  ATLAS detector}},  \href{http://xxx.lanl.gov/abs/1407.0583}{{\tt
  arXiv:1407.0583}}.

\bibitem{oxbridgeStransverseMassLibrary}
A.~J. Barr and C.~G. Lester, ``{Oxbridge Kinetics} / {Stransverse Mass
  Library}.''
\newblock \url{http://www.hep.phy.cam.ac.uk/~lester/mt2/index.html}.

\bibitem{Cheng:2008hk}
H.-C. Cheng and Z.~Han, {\it Minimal kinematic constraints and {$M_{T2}$}},
  {\em JHEP} {\bf 12} (2008) 063,
  [\href{http://xxx.lanl.gov/abs/0810.5178}{{\tt arXiv:0810.5178}}].

\bibitem{zenuhanStransverseMassLibrary}
H.-C. Cheng and Z.~Han, ``{UCD} stransverse mass library.''
\newblock
  \url{http://particle.physics.ucdavis.edu/hefti/projects/doku.php?id=wimpmass}.

\bibitem{Walker:2013uxa}
J.~W. Walker, {\it {A complete solution classification and unified algorithmic
  treatment for the one- and two-step asymmetric S-transverse mass $
  {\tilde{M}}_{\mathrm{T}2} $ event scale statistic}},  {\em JHEP} {\bf 1408}
  (2014) 155, [\href{http://xxx.lanl.gov/abs/1311.6219}{{\tt
  arXiv:1311.6219}}].

\bibitem{root}
R.~Brun and F.~Rademakers, {\it {ROOT - An Object Oriented Data Analysis
  Framework}},  {\em Nucl. Inst. and Meth. in Phys. Res. A} {\bf 389} (1997)
  81--86.

\bibitem{Etayo:2006:NAC:1159669.1159671}
F.~Etayo, L.~Gonzalez-Vega, and N.~del Rio, {\it A new approach to
  characterizing the relative position of two ellipses depending on one
  parameter},  {\em Comput. Aided Geom. Des.} {\bf 23} (May, 2006) 324--350.

\bibitem{HughesMohcine}
G.~Hughes and M.~Chraibi, {\it Calculating ellipse overlap areas},  {\em
  Computing and Visualization in Science} {\bf 15} (2012), no.~5 291--301.

\bibitem{Cho:2007dh}
W.~S. Cho, K.~Choi, Y.~G. Kim, and C.~B. Park, {\it {Measuring superparticle
  masses at hadron collider using the transverse mass kink}},  {\em JHEP} {\bf
  02} (2008) 035, [\href{http://xxx.lanl.gov/abs/0711.4526}{{\tt
  arXiv:0711.4526}}].

\end{thebibliography}\endgroup


\begin{table*}
\begin{center}
\begin{tabular}{ll|ccc}
  Method & & Precision (GeV) & Event Type & Time (s) \\
\hline
{\tt CH} & default & $\sim$ 0.002 & {\tt OVER} & 1.05 \\
{\tt LESTER(a) }& finite precision & \hphantom{$\sim$} 0.002 & {\tt OVER} & 0.59 \\
{\tt LESTER(b)} & finite precision & \hphantom{$\sim$} 0.002 & {\tt OVER} & 0.62 \\
{\tt LESTER(a) }& machine precision & $\sim$ $10^{-12}$  & {\tt OVER} & 1.86 \\
{\tt LESTER(b)} & machine precision & $\sim$ $10^{-12}$  & {\tt OVER} & 1.96 \\
{\tt CH} & default & $\sim$ 0.002  & {\tt UNDER} & 1.13 \\
{\tt LESTER(a) }& finite precision & \hphantom{$\sim$} 0.002 & {\tt UNDER} & 0.75 \\
{\tt LESTER(b)} & finite precision & \hphantom{$\sim$} 0.002 & {\tt UNDER} & 0.74 \\
{\tt LESTER(a) }& machine precision & $\sim$ $10^{-12}$  & {\tt UNDER} & 2.16 \\
{\tt LESTER(b)} & machine precision & $\sim$ $10^{-12}$  & {\tt UNDER} & 2.22 \\
\end{tabular}

  \caption{
    This table shows the time in seconds taken to perform 400,000 $\mttwo$ evaluations using different implementations, each on two sorts of event (one for which {\tt CH} tended to over-estimate $\mttwo$ values, `{\tt OVER}',  and one for which {\tt CH} tended to under-estimate $\mttwo$ values `{\tt UNDER}').  \changed{All programs were compiled using the {\tt clang-600.0.56} C-compiler, with {\tt -O3} optimisations turned on, and were run on a 1.4 GHz Intel Core i5 MacBook air running OSX version 10.10.1. }
    The three main methods compared are (i) the old {\tt CH} algorithm (known to be accurate to about 0.002 GeV), (ii) the new algorithm `{\tt LESTER}' running in machine precision mode, and (iii) the new algorithm `{\tt LESTER}' running to a lesser precision of 0.002 GeV.  \changed{The new algorithm is demonstrated running in two variants: {\tt LESTER(a)} and {\tt LESTER(b)}, the former being the default implementation as described in the text (i.e.~with the deci-section optimization described in Section~\ref{sec:decisection} turned on) and the latter being almost the same but with the deci-section optimsation turned off.} It may be seen that the new algorithm is no worse than 20\% to 50\% slower than {\tt CH} when working in {\em machine-precision mode} (i.e.~generating much better answers than {\tt CH}).  \changed{It is evident that when asked to only attempt
    to get answers to 0.002 GeV, the new algorithm was then about two to three times faster than the old {\tt CH} algorithm at similar precision.  The `{\tt OVER}' events consist of two 175 GeV top quarks and a jet being produced approximately 200 GeV above threshold, with each top quark decaying to a on-shell $W$ boson (leptonically decaying) and an unhadronized $b$-quark.
    The visible systems fed to $\mttwo$ were each the sum of the momenta (without combinatoric uncertainty) of the $b$-quark and lepton from one of the tops, the missing transverse momentum was the sum of the two neutrino monenta,  all momenta were unsmeared montecarlo truth, $\chi$ was taken to be zero, and leptons, $b$-jets and neutrinos were taken to be massless.
    The `{\tt UNDER}' events consit of two 300 GeV sleptons and an ISR jet being produced approximately 200 GeV above threshold, with each slepton subsequently decaying to a 0.001 GeV neutralino and a 0.001 GeV lepton.  The visible systems fed to $\mttwo$ were lepton momenta, the missing transverse momentum was the sum of the two neutralino monenta,  all momenta were unsmeared montecarlo truth, $\chi$ was taken to be
  0.001 GeV.}
    \label{tab:timing}
  }
\end{center}
\end{table*}


\end{multicols}
\end{document}